# Periodic-orbit analysis and scaling laws of intermingled basins of attraction in an ecological dynamical system

R. F. Pereira [1], S. Camargo [2], S. E. de S. Pinto [3], S. R. Lopes [1], and R. L. Viana [1] *

1. *Departamento de Física, Universidade Federal do Paraná,*

*81531-990, Curitiba, Paraná, Brazil.*

2. *Instituto de Física, Universidade de São Paulo,*

*05315-970, São Paulo, São Paulo, Brazil.*

3. *Departamento de Física, Universidade Estadual de Ponta Grossa,*

*84032-900, Ponta Grossa, Paraná, Brazil*

(Dated: September 29, 2008)

## Abstract

Chaotic dynamical systems with two or more attractors lying on invariant subspaces may, provided certain mathematical conditions are fulfilled, exhibit intermingled basins of attraction: each basin is riddled with holes belonging to basins of the other attractors. In order to investigate the occurrence of such phenomenon in dynamical systems of ecological interest (two-species competition with extinction) we have characterized quantitatively the intermingled basins using periodic-orbit theory and scaling laws. The latter results agree with a theoretical prediction from a stochastic model, and also with an exact result for the scaling exponent we derived for the specific class of models investigated. We discuss the consequences of the scaling laws in terms of the predictability of a final state (extinction of either species) in an ecological experiment.

---

* Corresponding author. e-mail: viana@fisica.ufpr.br

## I. INTRODUCTION

Experiments on the competition of two species of flour beetles, *Tribolium castaneum* and *Tribolium confusum* resulted in the eventual extinction of either one of the two competitors [1]. However, the particular species to become extinct was found to be extremely sensitive to the initial population and the environmental conditions (such as temperature and humidity) prevailing during the realization of the experiment [2]. This competitive indeterminacy problem has been extensively investigated: in replicate cultures of *T. castaneum* and *T. confusum* one species exterminates the other with a probability $p$ inasmuch for other cultures, the other species wins with probability $1 - p$ [3].

These features are traditionally explained by means of two kind of hypotheses. The first, called *genetic stochasticity*, attributes the identity of the winning species mainly to the genetic characteristics of the founding populations [3]. This means that, if some species eventually wins, a genetically superior collection of beetles of this species is matched with a relatively inferior population of beetles of the opposite species, and vice-versa. The competitive indeterminacy will be due to random sampling among the genotypes distributed from a stock culture [4]. The second hypothesis, called *demographic stochasticity*, is the variability in the population growth rates arising from random differences among individuals in survival and reproduction within a season. This variability happens even if all individuals have the same expected ability to survive and reproduce, and if the expected rates of survival and reproduction do not change from one generation to the next [3].

Besides these conventional theories based on some kind of stochastic behavior, that give a satisfactory explanation to the experiments, there is also an alternative approach based on a different type of competitive indeterminacy caused by deterministic factors, and that will be followed in the present paper. The starting point is the existence of experimental evidences of chaotic behavior in the time evolution of the populations of single species of *Tribolium* [5]. Since the time evolution of this simple two-species competition system is thought to be deterministic, we may regard the extinction of either species as an asymptotic state with a well-defined attractor in the phase space [6]. The observed extreme sensitivity on the initial condition has led to the hypothesis that the basins of these attractors exhibit an extreme form of fractality called riddling [7].

A dynamical system is said to present riddled basins when it has a chaotic attractor

A whose basin of attraction is riddled with holes belonging to the basin of another (non necessarily chaotic) attractor B [8, 9]. Riddling means that every point in the basin of attractor A has pieces of the basin of attractor B arbitrarily nearby. A consequence of riddling is that, no matter how small is the uncertainty with which an initial condition is determined, the asymptotic state of the system remains virtually unpredictable, thus defying the formally deterministic character of the model [10, 11].

Riddled basins of attraction have been observed both in mathematical models [8–11] as well as in experiments [12]. In particular, riddling has been described in ecological population models [13]. The basins are called *intermingled*, when each basin is riddled with holes belonging to the other basin. In this case there must be at least two attractors lying in different invariant subspaces, and the basin of each attractor is pierced with holes containing initial conditions belonging to the basin of the other attractor [8, 14]. Moreover, these basins are so intertwined that, given an initial condition with a finite uncertainty, the final state cannot be predicted. The sensitivity observed in the *Tribolium sp.* experiments suggested that a mathematical model describing the problem exhibits intermingled basins, since the initial population in the experiments correspond to an initial condition which is unavoidably plagued with some uncertainty, and thus with uncertain outcome if the basins are intermingled [6, 7]

For intermingled basins to exist there must be chaotic orbits in the invariant subspaces of the two coexisting attractors. The natural ergodic measure in each attractor is supported by an infinite number of unstable periodic orbits, and a quantitative characterization of the chaotic orbits is thus furnished by a periodic orbit analysis. Commonly used diagnostic in the latter are the finite-time Lyapunov exponents and the contrast measure [15, 16]. However, in order to relate the presence of intermingled basins to some mathematical model of the competition between the *Tribolium* species, it is necessary to explore some quantitative consequences of this basin property, particularly in the form of scaling laws that can be, at least in principle, confirmed by future experiments. Such scaling laws have been derived by Ott and coworkers from a stochastic model with drift and reflecting barrier [17].

In this paper we consider a possible mathematical model of the *Tribolium* experiments based on the previous works of Hofbauer *et al.* [6] and Kan [7], and which presents intermingled basins. We verify the scaling laws for riddled basins and compare the numerical results with the above-mentioned theoretical model [17]. These scaling laws for the intermingled basins represent quantitative characterizations that can be used to establish connections between the existence of such basins in this and other related experiments (the existence of locally intermingled basins has been reported in a system of two coupled Chua circuits [18]). Moreover, a periodic orbit analysis is performed so as to investigate the role of the unstable periodic orbits in the metric properties of the chaotic orbits in the system, since the presence of transversely unstable periodic orbits is the dynamical phenomenon that enables the existence of intermingled basins.

The rest of this paper is organized as follows. In Section II we present the mathematical model for the *Tribolium sp.* experiment as well some of its mathematical properties. Section III deals with the characterization of intermingled basins in this model using typical chaotic orbits in the corresponding attractors. The same characterization, but now using periodic orbit analysis (i.e., unstable periodic orbits embedded in the attractors) is the object of Section IV. The scaling laws that can be used to identify and investigate intermingled basins are the subject of Section V: the fraction of basin areas close to the attractors and the uncertain fraction of initial conditions. The last section is devoted to our conclusions.

## II. INTERMINGLED BASINS IN A TWO-SPECIES COMPETITION MODEL

The red flour beetles *Tribolium castaneum* and *Tribolium confusum* have similar habitats and identification. Both primarily attack milled grain products, as flour and cereals. They are a very common insect pest of flour mills in many countries, since they are particularly injurious in warehouses and in factories making starch products. Experiments involving the evolution of these two competing species have shown as the final outcome the extinction of either one of them. What species becomes extinct seem to depend on the initial population and environmental conditions in an extremely sensitive manner [1].

Hofbauer and coworkers [6] have developed a class of two-dimensional models in which we consider two species with populations $x_1(n)$ and $x_2(n)$ at (discrete) times $n = 0, 1, 2, \ldots$, which labels the insect generation, and satisfying identical evolution equations

$$x_1(n+1) = x_1(n)\Phi(x_1(n) + x_2(n)), \quad (1)$$
$$x_2(n+1) = x_2(n)\Phi(x_1(n) + x_2(n)), \quad (2)$$

where $\Phi(.)$ is a function compatible with three biological requirements: (i) the proportion of each species does not change with time, i.e. $x_2(n+1)/x_1(n+1) = x_2(n)/x_1(n)$; (ii) the total population $x = x_1 + x_2$ in a generation depends only on the value at its previous generation; and (iii) $\partial\Phi/\partial x_1 < 0$ and $\partial\Phi/\partial x_2 < 0$.

In order to exhibit competition and species extinction, this model includes perturbations of Eqs. (1)-(2) in the general form

$$x_1(n+1) = x_1(n)\Phi(x_1(n) + x_2(n))\left[1 + \kappa x_1(n)G(x_1(n), x_2(n))\right], \tag{3}$$

$$x_2(n+1) = x_2(n)\Phi(x_1(n) + x_2(n))\left[1 - \kappa x_2(n)G(x_1(n), x_2(n))\right], \tag{4}$$

where $0 < \kappa < 1$ stands for the strength of the competition between species and, for simplicity, we assume that $G(.)$ depends only on the total population $x = x_1 + x_2$. Changing variables from $(x_1, x_2)$ to $(x, y \equiv x_1/x)$ we obtain, from Eqs. (3)-(4), the following two-dimensional map

$$x(n+1) = T(x(n)) \equiv x(n)\Phi(x(n)), \tag{5}$$

$$y(n+1) = y(n) + \kappa y(n)(1 - y(n))g(x(n)), \tag{6}$$

where $g(x) = xG(xy, x(1-y))$ satisfies the mathematical requirements stated in Ref. [6].

Within the class of two-dimensional discrete models defined by Eqs. (5)-(6) both functions $T$ (or $\Phi$) and $g$ (or $G$) have to be determined taking into account the possibility of riddling basins. There must be two possible outcomes: $y = 0$ (extinction of the $x_1$ species) or $y = 1$ (extinction of the $x_2$ species). Those results show up as the only attractors of the two-dimensional map in the phase plane, denoted as A and B, respectively. Moreover, their basins of attractions, denoted as $\beta(\mathsf{A})$ and $\beta(\mathsf{B})$, must be riddled (in fact intermingled) what poses additional requirements in the formulation of the model.

Let us briefly review the set of conditions defining riddled basins for two-dimensional maps [10]. When the basin of attraction of A is riddled with holes belonging to the basin of another attractor B, we can say that, if a randomly chosen point has a positive probability of being in $\beta(\mathsf{A})$, then it also has positive probability of not being in $\beta(\mathsf{A})$. In the latter case, the point belongs to the other basin of attraction $\beta(\mathsf{B})$.

This measure-theoretical definition and the fact that one basin is riddled by the other one, implies the following set of conditions under which riddled basins occur in such a two-dimensional dynamical system [10]:

1. there are two invariant one-dimensional subspaces $\mathsf{I}_0$ and $\mathsf{I}_1$ in the phase plane;

2. the dynamics on the invariant subspaces $I_0$ and $I_1$ have chaotic attractors A and B, respectively;

3. the attractors A and B are transversely stable in the phase plane, i.e. for typical orbits on the attractors the Lyapunov exponent for infinitesimal perturbations along the direction transversal to the invariant subspaces $I_0$ and $I_1$, respectively, is negative;

4. a set of unstable periodic orbits embedded in the chaotic attractors A and B are transversely unstable. As a consequence, along the direction transversal to $I_0$ and $I_1$, the Lyapunov exponent experiences positive finite-time fluctuations.

Condition **1** is a consequence of the system having some symmetry which enables it to display invariant subspaces $I_i$, $i = 0, 1$ in the sense that, once an initial condition is exactly placed on $I_i$, the resulting trajectory cannot escape from $I_i$ for further times. To have riddling, (and thus intermingled basins) it is necessary to exist dense sets of points with zero Lebesgue measure in the attractors A and B lying in the invariant subspaces $I_0$ and $I_1$ which are transversely unstable, thus it is necessary that these attractors be chaotic, what is the content of condition **2**.

If the transverse Lyapunov exponent of typical orbits lying in the invariant subspaces $I_0$ and $I_1$ is negative (condition **3**), then A and B are attractors at least in the weak Milnor sense, and their basins have positive Lebesgue measure. Condition **4** states that, while the invariant subspaces $I_i$ are still transversely stable, there will be (atypical) trajectories on the attractors A and B that are transversely unstable. Condition **3** can be quantitatively checked by computing the Lyapunov exponent along the transversal direction to $I_i$. Verifying condition **4**, on the other hand, would require the determination of a transversely unstable periodic orbit embedded in the attractor A or B. It turns out that this is feasible only for a few dynamical systems [19]. In most situations we shall resort to other ways to verify the existence of finite-time fluctuations by computing the finite-time Lyapunov exponents.

The evolution of the total population, governed by the one-dimensional map $T(x)$, must be chaotic in order to fulfill condition **2** for riddling, since the only way to have an infinite number of unstable periodic orbits within an attractor is to ensure the existence of a dense chaotic orbit lying in the invariant manifold. The transversal dynamics (6) has two invariant subspaces: $I_0 = \{(x, y = 0) | x \in [0, 1]\}$ and $I_1 = \{(x, y = 1) | x \in [0, 1]\}$, what fulfills condition **1**. In addition, although this is not a necessary condition, we suppose that the attractors

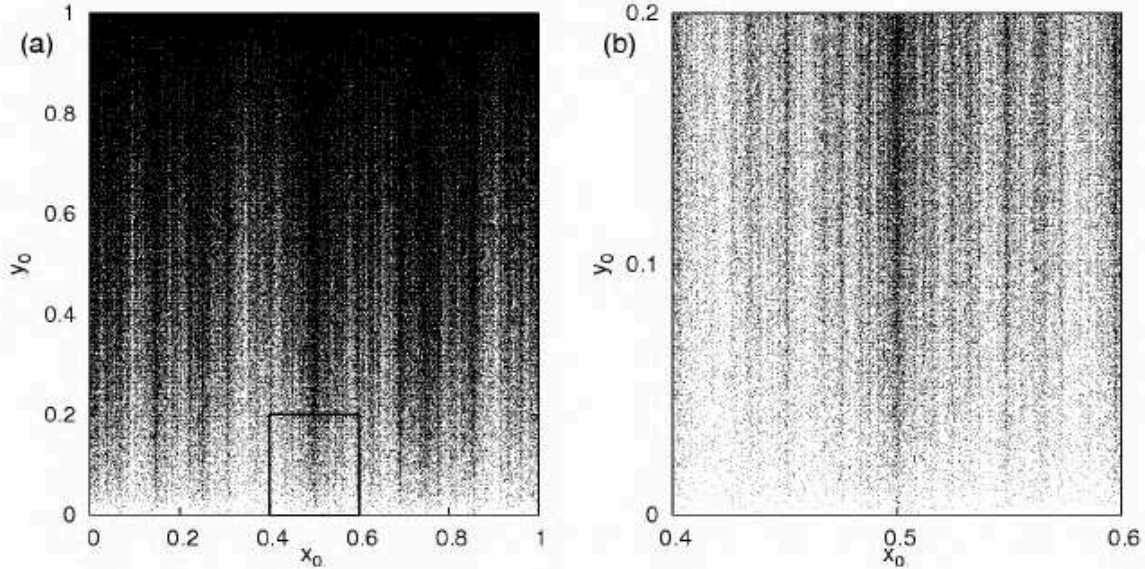

FIG. 1: Basins of the attractor at $I_0 : y = 0$ (white pixels) and $I_1 : y = 1$ (black pixels) when $\kappa = 0.3$. (b) Magnification of a small region of (a).

A and B have measures occupying the whole intervals $I_0$ and $I_1$, respectively. The specific function $g(x)$, on the other hand, must be chosen so as to warrant the proper transverse stability conditions **3** and **4**.

The case for which $T(x) = 3x \pmod 1$ and $g(x) = \cos(2\pi x)$ was previously studied by Kan [7], who proved the existence of intermingled basins when $\kappa = 1/32$. This proof has been extended to the case $0 < \kappa < 1$ by Hofbauer and collaborators [6]. In the present paper we consider a slightly modified version of Kan's model, by choosing

$$x(n+1) = 4x(n)(1-x(n)), \tag{7}$$
$$y(n+1) = y(n) + \kappa y(n)(1-y(n))\cos(3\pi x(n)). \tag{8}$$

such that the dynamics in each invariant subspace is strongly chaotic, with natural measure $d\mu(x) = \left(\pi\sqrt{x(1-x)}\right)^{-1} dx$.

A representative example of the basins of attraction exhibited by this system is depicted in Figure 1(a) for $\kappa = 0.3$. A numerical approximation of the basin of the attractor at $I_0$ ($I_1$) is represented by the white (black) pixels, and show an intertwined structure, with pieces of a basin approaching the other attractor at arbitrarily small distances. Moreover, there is a tongue-like structure of each basin with self-similar character, as suggested by the magnification shown in Fig. 1(b).

Two basic properties of the intermingled basins are apparent in Fig. 1: (i) there are pieces of some basin at distances arbitrarily close to the other attractor; (ii) the vicinity of a point belonging to any basin contains points belonging to the other basins, at small scales. These observations are to be confirmed and quantified by derivation of two scaling laws involving information from the finite-time Lyapunov exponents. Actually these properties follow from the tongue-like structure of the basin filaments next to the subspaes. The tongues are anchored at transversely unstable periodic orbits belonging to each subspace. Since these orbits form a dense set, the tongue-like structure is likewise dense.

## III. CHARACTERIZATION OF INTERMINGLED BASINS BY TYPICAL CHAOTIC ORBITS

In order to discuss conditions **3** and **4** for riddled basins in a quantitative setting, it is useful to work with the finite-time Lyapunov exponents of the two-dimensional map defined by Eqs. (7)-(8). Let $n$ be a positive integer and $\mathbf{DF}^n(x(0), y(0))$ be the Jacobian matrix of the $n$ times iterated map $\mathbf{F}(x,y) = (4x(1-x), y + \kappa y(1-y)\cos(3\pi x))$, with entries evaluated at an initial condition $(x(0), y(0))$. We assume that the singular values of $\mathbf{DF}^n(x(0), y(0))$ are ordered as : $\xi_1(x(0), y(0), n) \geq \xi_2(x(0), y(0), n)$. Then, the time-$n$ Lyapunov exponents for the point $x(0), y(0)$ are defined as [20]

$$\tilde{\lambda}_1(x(0), y(0); n) = \frac{1}{n} \ln ||\mathbf{DF}^n(x(0), y(0)).\mathbf{v}_1||, \qquad (9)$$

$$\tilde{\lambda}_2(x(0), y(0); n) = \frac{1}{n} \ln ||\mathbf{DF}^n(x(0), y(0)).\mathbf{v}_2||, \qquad (10)$$

where $\mathbf{v}_{1,2}$ is the singular vector related to $\xi_{1,2}(x(0), y(0), n)$, respectively.

Due to the skew-product structure of Eqs. (7)-(8), as well as of the whole class of maps described by Hofbauer et al. [6], the dynamics at direction 1 is not affected by the dynamics at direction 2 (the converse, however, is obviously true). Hence we may fix $\xi_1 = \xi_{||}$ as being the eigenvalue at directions parallel to the invariant subspaces $I_{0,1}$, with $\xi_2 = \xi_{\perp}$ denoting the transverse direction.

The infinite time-limit of Eqs. (9)-(10) are the usual Lyapunov exponents:

$$\lambda_{\|} = \lim_{n\to\infty} \tilde{\lambda}_{\|}(x(0), y(0), n) = \int_0^1 \frac{4(1-2x)dx}{\pi\sqrt{x(1-x)}} = \ln 2, \qquad (11)$$

$$\lambda_{\perp} = \lim_{n\to\infty} \tilde{\lambda}_{\perp}(x(0), y(0), n) = \lim_{n\to\infty} \frac{1}{n} \sum_{i=1}^{n} \ln ||\mathbf{DF}(x_i, y_i).\mathbf{e}_y||, \qquad (12)$$

where the first result comes from considering *typical* chaotic trajectories in the attractors A or B, for which Birkhoff's ergodic theorem can be applied. Moreover, although the time-$n$ exponent $\tilde{\lambda}_k(x(0), y(0), n)$, $k = 1, 2$, has in general a different value, depending on the point we choose, its infinite time limit takes on the same value for almost all $(x(0), y(0))$ with respect to the natural ergodic measure of the invariant sets A or B.

If an attractor has a riddled basin, it must be transversely stable (condition **3** for riddling), such that it is necessary that $\lambda_{\perp} < 0$. On the other hand, it is also required that the attractor contains transversely unstable orbits (condition **4**). This is possible because in a transversely stable chaotic attractor there can be an infinite number of transversely unstable orbits. This implies the existence of positive and negative fluctuations of the finite-time transversal exponent, $\tilde{\lambda}_{\perp}$, what makes useful to work with the probability distribution $P(\tilde{\lambda}_{\perp}(x(0), y(0), n))$, from which we can obtain the average value of this exponent (assuming proper normalization):

$$\left\langle \tilde{\lambda}_{\perp}(x(0), y(0), n) \right\rangle = \int_{-\infty}^{+\infty} \tilde{\lambda}_{\perp}(x(0), y(0), n) P(\tilde{\lambda}_{\perp}(x(0), y(0), n)) d\tilde{\lambda}_{\perp}(n). \qquad (13)$$

When $n$ is large enough the form of this distribution can be written as [20]

$$P(\tilde{\lambda}_{\perp}(x(0), y(0), n)) \approx \sqrt{\frac{nG''(\lambda_{\perp})}{2\pi}} e^{-nG(\tilde{\lambda}_{\perp})}, \qquad (14)$$

where the function $G(\lambda)$ has the following convexity properties:

$$G(\lambda_{\perp}) = G'(\lambda_{\perp}) = 0, \qquad G''(\lambda_{\perp}) > 0. \qquad (15)$$

Expanding $G(\lambda)$ in the vicinity of $\lambda_{\perp}$, the first non-vanishing term is the quadratic one, such that $P(\tilde{\lambda}_{\perp}(x(0), y(0), n))$ reduces to a Gaussian distribution

$$P_{\perp}(\tilde{\lambda}_{\perp}(n)) \approx \sqrt{\frac{nG''(\lambda_{\perp})}{2\pi}} \exp\left[-\frac{nG''(\lambda_{\perp})}{2}(\tilde{\lambda}_{\perp}(n) - \lambda_{\perp})^2\right], \qquad (n \gg 1) \qquad (16)$$

such that, on substituting into (13) there results

$$\left\langle \tilde{\lambda}_{\perp}(x(0), y(0), n) \right\rangle = \lambda_{\perp}. \qquad (17)$$

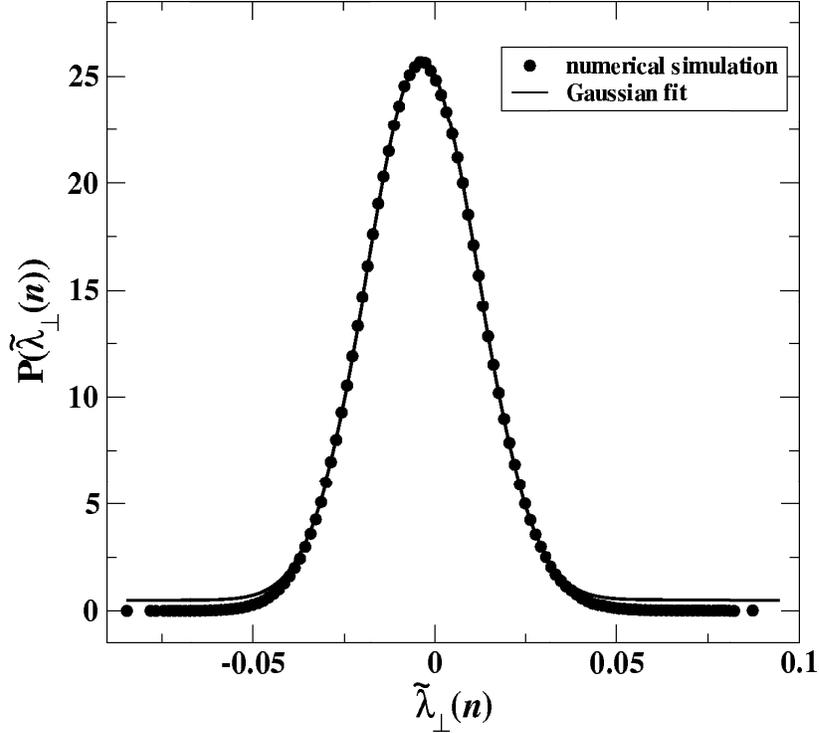

FIG. 2: Probability distribution function $P(\tilde{\lambda}_\perp)(n)$ for the time-24 transverse Lyapunov exponents of attractor A, with $\kappa = 0.1$.

Since the standard deviation in the Gaussian approximation approaches zero with $n^{-1/2}$, when $n \gg 1$, we can obtain the corresponding variance as

$$\sigma^2 = <\left(\tilde{\lambda}_\perp(n) - <\lambda_\perp>\right)^2>. \tag{18}$$

We show in Fig. 2 the numerically obtained probability distribution function (PDF) $P(\tilde{\lambda}_\perp)(x(0), y(0) = 1, n)$ for the time-24 transverse Lyapunov exponents of the attractor A when $\kappa = 0.1$, and compare it with a Gaussian fit given by Eq. (16). The latter gives better results for the bulk of the distribution, with small deviation in its tails. Hence, on supposing a Gaussian diffusive process, we can define a diffusion coefficient

$$D = \frac{2\sigma^2}{n} = \frac{1}{2} G''(\lambda_\perp), \tag{19}$$

where we have used Eq. (14) [22].

In Figure 3(a) we depict (in gray-scale) the dependence on the parameter $\kappa$ of the numerically obtained PDF $P(\tilde{\lambda}_\perp(x(0), y(0) = 1, n))$ for the time-24 transverse Lyapunov exponents of the attractor $\mathbf{A}$. The PDFs drift toward negative values as $\kappa$ increases, hence their average values are negative for any $\kappa > 0$: $\lambda_\perp < 0$ (condition **3**). The width of these PDFs are nearly constant with respect to $\kappa$, with a degree of asymmetry varying from negative to positive values, as shown by Fig. 3(b), where some of the moments of the PDFs are depicted as a function of $\kappa$. These results were obtained by using typical chaotic orbits, but the observed behavior is supported by the structure of unstable periodic orbits embedded in both attractors [21].

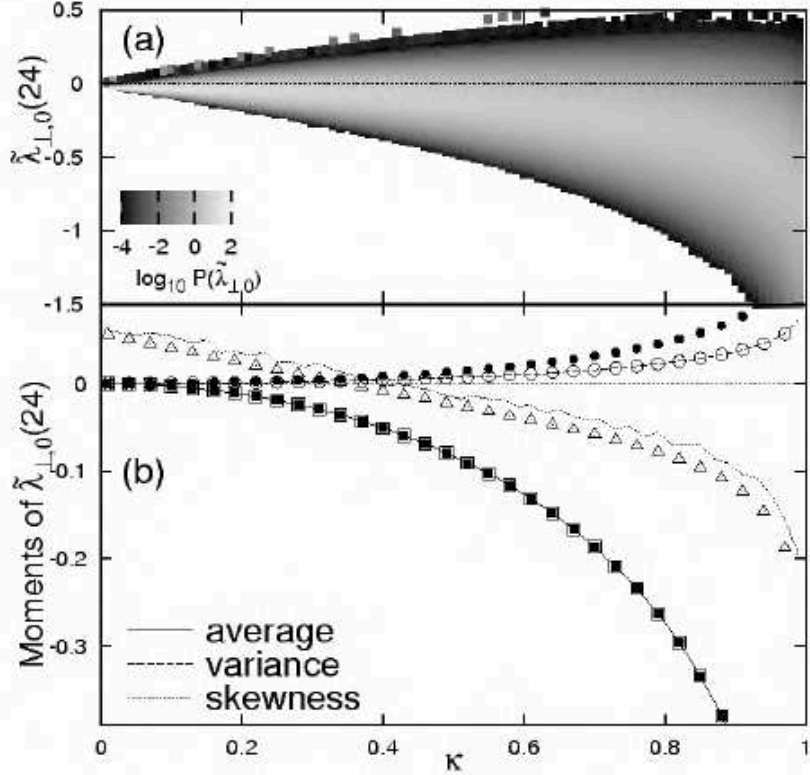

FIG. 3: (a) Probability distribution function (in gray-scale) for time-24 transversal Lyapunov exponents for typical orbits in the subspaces $y = 0$ as a function of the parameter $\kappa$. (b) Moments of the probability distribution function *versus* $\kappa$. The lines stand for typical chaotic orbits, whereas the symbols represent atypical unstable period-$p$ orbits (open symbols of $p = 24$, filled symbols for $p = 12$).

The existence of transversely unstable orbits in the attractors (condition **4**) implies that there is a positive fraction of positive values of $\tilde{\lambda}_\perp(n)$ for initial conditions $(x(0), y(0))$

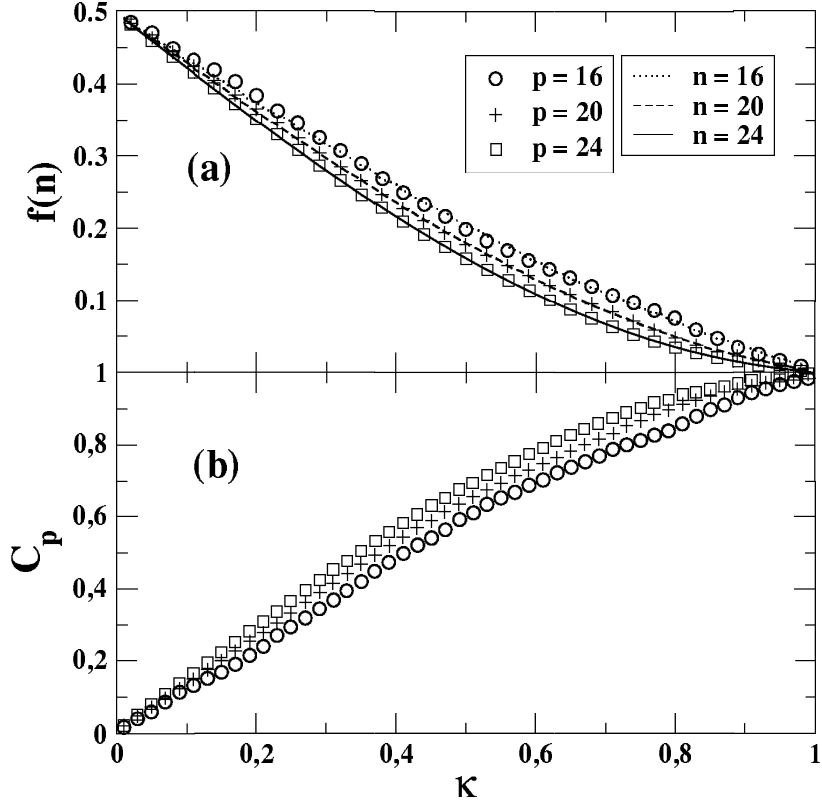

FIG. 4: Dependence with $\kappa$ of (a) the positive fraction of transversal Lyapunov exponents for typical orbits ($n = 16, 20, 24$) and unstable periodic orbits ($p = 16, 20, 24$); (b) contrast measure for $p = 16, 20, 24$. The orbits have been chosen from the attractor at $I_0 : y = 0$, but identical results were found for the other attractor at $I_1 : y = 1$.

randomly chosen in the attractor A or B:

$$f(n) = \int_0^\infty P(\tilde{\lambda}_\perp(x(0), y(0), n)) d\tilde{\lambda}_\perp(n) > 0. \tag{20}$$

If the distribution of finite-time exponents is symmetric and so as to have half of their values with positive sign, i.e., $f(n) = 1/2$, then the infinite-time exponent $\lambda_\perp$ vanishes, and the attractor loses transversal stability, configuring a *blowout bifurcation* [23]. We remark that the occurrence of a blowout bifurcation marks the endpoint of riddling, since after that the invariant chaotic sets become transversely unstable.

The dependence of the positive fraction of time-$n$ transverse Lyapunov exponents with $\kappa$ is depicted in Fig. 4(a) for three different values of $n$. The results indicate that, for $0 < \kappa < 1$ we have a nonzero positive fraction of transversal exponents, on account of the existence of transversely unstable periodic orbits. Since the infinite-time Lyapunov exponent is already negative for this range of parameters there follows that the map always fulfill the conditions for riddling. This result holds for both attractors A and B, such that their basins of attraction are intermingled.

As $\kappa$ approaches the unity, the positive fraction goes to zero. In this case the populations keep their proportions unchanged, such that there is no extinction of any population. This marks the onset of a phenomenon called unstable dimension variability (UDV), which underlies the indeterminacy characteristic of the species extinction. A chaotic invariant set possessing UDV has, embedded on it, periodic orbits with a different number of unstable directions [24]. A particularly troublesome consequence of UDV is the lack of adequate shadowability properties of noisy trajectories, such as those obtained by using computers, where the role of noise is played by unavoidable one-step roundoff and truncation errors [25].

If UDV is too severe, it may happen that a noisy trajectory is not closely followed by *any* fiducial trajectory of the original system for a reasonable time. Hence, the computer-generated trajectories in this case may be just numerical artifacts, and no relevant statistics can be extracted from such orbits [26]. In this case, even though the system is formally a deterministic one, the character of the orbits is, at best, of a stochastic system. In fact, when there is no shadowability at all, the mathematical model itself may be of limited use, and one should resort to experimental data (using embedding techniques, for example) to obtain relevant information about the system dynamics. UDV was first described for a diffeomorphism in $T^2 \times S^2$ [27]. The earliest observation of UDV for a dynamical system of physical interest was reported for the kicked double rotor map [28, 29]. The presence of UDV seems to be typical in high-dimensional dynamical systems, as in coupled map lattices [30]. The analysis developed in this Subsection for the *Tribolium* system system shows immediately that it exhibits UDV for all values of the parameter $0 < \kappa < 1$.

## IV. CHARACTERIZATION OF INTERMINGLED BASINS BY UNSTABLE PERIODIC ORBITS

The natural measure of a chaotic attractor, generated by a *typical* chaotic orbit, is supported by an infinite number of unstable periodic orbits embedded in that attractor. The latter are thus *atypical* orbits. Each unstable orbit has thus its own measure with a well-defined contribution to the natural measure. The object of periodic orbit theory is to characterize this contribution in order to elucidate the underlying mechanisms involving global bifurcations like the onset of riddling [21].

For ease of notation we write $\mathbf{x}_i = (x_i, y_i = 0, 1)$ and, accordingly, denote as $\{\mathbf{x}_1(j), \mathbf{x}_2(j), \cdots, \mathbf{x}_p(j)\}$ the points of a $j$th period-$p$ orbit embedded in the chaotic attractor, where $j = 1, 2, \cdots N_p$, with $N_p$ as the total number of period-$p$ orbits. In this case $\mathbf{x}_p(j)$ is the $j$th fixed point of $\mathbf{F}^p(\mathbf{x})$, i.e., $\mathbf{x}_p(j)$ is on a period-$r$ orbit, where $r$ equals either $p$ or a prime factor of $p$. We can define a transverse Lyapunov exponent for such period-$p$ orbit as [15]

$$\lambda_\perp(\mathbf{x}_p(j), p) = \frac{1}{p} \sum_{i=1}^{p} \ln ||\mathbf{DF}(\mathbf{x}_i(j)).\mathbf{e}_y||, \qquad (21)$$

in such a way that a typical chaotic orbit on either attractor has its natural measure supported by transversely stable and unstable periodic orbits, i.e. those for which $\lambda_\perp(p, j)$ is negative and positive, respectively.

We would expect that, the higher the period-$p$ the most likely the atypical measure of the unstable orbit would approach the typical nature measure of the chaotic orbit. This is confirmed by the numerical results summarized in Figure 3(b). The number of period-$p$ orbits is known to increase exponentially with $p$, according to the topological entropy of the attractors (in the case of the *Tribolium* system we have $h_T = \ln 2$ for both attractors), thus reliable statistical results can only be obtained with large periods $p$.

Some moments of the probability distribution function for unstable orbits of periods $p = 12$ and $24$ are depicted in Figure 3(b) as a function of $\kappa$, and they agree with the result obtained for the Lyapunov exponents computed from typical chaotic orbits. Likewise we can compute the positive fraction $f(n)$ of transversal Lyapunov exponents for orbits with different periods, and Fig. 4(a) shows results in agreement with those obtained from typical orbits. By construction, the relative number of transversely unstable orbits is equal to the positive fraction $f(n)$ given by Eq. (20).

Let us denote by $\mu_p(j)$ the natural measure of a typical trajectory in the neighborhood of the $j$th period-$p$ orbit. The measure $\mu_p(j)$ may be regarded as the probability that a typical chaotic trajectory in the attractor visits the neighborhood of the periodic orbit $\mathbf{x}_p(j)$. This probability is as smaller as the more unstable the periodic orbit, such that orbits with large unstable eigenvalues have a comparatively smaller contribution to the natural measure.

The (atypical) natural measure associated with the $j$th period-$p$ orbit is the normalized ratio [16]

$$\mu_p(j) = \frac{1/L_u(\mathbf{x}_p(j))}{\sum_{\ell=1}^{N_p}[1/L_u(\mathbf{x}_p(\ell))]}, \tag{22}$$

where $L_u$ is the absolute value of the product of eigenvalues related to the unstable directions of $\mathbf{DF}^p$, computed for the orbit points $\mathbf{x}_p(j)$. The unstable direction along the invariant subspace has usually the greater eigenvalue. Summing up over all unstable period-$p$ orbits embedded in the attractor (**A** or **B**) gives then its natural measure when the period goes to infinity [31]

$$\mu = \lim_{p \to \infty} \sum_j \mu_p(j), \tag{23}$$

It is possible to quantify the relative contribution of the period-$p$ orbits to the natural measure of a chaotic attractor by defining the contrast measure [32],

$$C_p = |\mu_p^u - \mu_p^s|, \tag{24}$$

where

$$\mu_p^s = \sum_{i=1}^{N_p^s} \mu_p(i), \quad \mu_p^u = \sum_{i=1}^{N_p^u} \mu_p(i), \tag{25}$$

and $\mu_p(i)$ is given by Eq. (22). The contrast measure is depicted in Figure 4(b) as a function of the parameter $\kappa$ for orbits with three different periods. The results confirm those obtained with help of finite-time Lyapunov exponents [Fig. 4(b)]: the system exhibits riddling (and intermingled basins) and unstable dimension variability for all values of $\kappa$.

## V. SCALING LAWS FOR RIDDLED BASINS

In the previous section we have used finite-time Lyapounov exponents and periodic-orbit theory to demonstrate the existence of intermingled basins for our dynamical system for all values of its control parameter $\kappa$. Since it is intended to be a model of an ecological experiment, we would be also interested to verify scaling relations characteristic of riddled basins, and that are more amenable no numerical analysis. Such scaling laws have been derived by Ott and co-workers and can be theoretically explained, at least within a specified parameter interval, by a stochastic model consisting on a biased random walk with reflecting barrier. The parameters of the latter model are drawn from the properties of the finite-time Lyapunov exponents that we have presented in the previous Section.

## A. Fraction of basin areas in the neighborhood of an attractor

Let us consider the phase portrait of Fig. 1 showing the intermingled basins of the two attractors, focusing on a horizontal line at $y = y_0$, which is the distance to the attractor A at $I_0$. We evaluate the fraction of its length that belongs to the basin of the other attractor B at $I_1$. This fraction will be denoted $P_1(y_0)$, is the probability of a trajectory starting at a distance $y_0$ from the attractor at $I_0$ to belong to the basin of the attractor at $I_1$.

If the basin of A is riddled with tongues belonging to the basin of B, there follows that for any distance $y_0$, no matter how small, there is always a nonzero value of $P_1(y_0)$. Since the pieces of the basin of B are anchored at the invariant manifold $y = 0$, this fraction tends to zero as $y_0 \to 0$ (in the limit it is a Lebesgue measure zero set), and is expected to scale with $y_0$ as a power-law

$$P_1(y_0) \sim |y_0|^\eta, \tag{26}$$

where $\eta > 0$ is a characteristic scaling exponent. Our numerical results are shown in Fig. 5(a) for three different values of $\kappa$, confirming the validity of the scaling law (26) with an exponent $\eta \approx 1$. Figure 5(b) presents the variation of the numerically obtained scaling exponent with $\kappa$ (filled squares).

The stochastic model of Ref. [10, 17] predicts for this scaling exponent the following value [10, 17]:

$$\eta = \frac{|\lambda_\perp|}{D}, \tag{27}$$

where $D$ is the diffusion coefficient given by Eq. (19), which was obtained through the variance of the finite-time exponent fluctuations. Figure 5(b) shows that the values obtained from this stochastic model roughly agree with the numerical values but, even so, fails for large $\kappa$. The reason for this disagreement is that the stochastic model is expected to work better when the process approach most a random walk, where are small correlations between the points. The random-walk limit is, strictly speaking, the blowout bifurcation point. The reason for this lies in the periodic-orbit theory: at the blowout point the infinite-time transverse Lyapunov exponent vanishes, meaning that the contribution of the transversely stable orbits embedded in the chaotic attractor counterbalances the contribution of the transversely unstable orbits. Hence, at the blowout point, an orbit in the vicinity of the chaotic attractor is subjected to an equal number of repelling and attracting contribution from the unstable orbits, what is equivalent to a random walk in statistical terms (although the orbit points never cease to have some correlation). We stress that, although $\kappa = 0$ is a blowout point, this does not configure a blowout bifurcation, since we have restricted $\kappa$ to non-negative values.

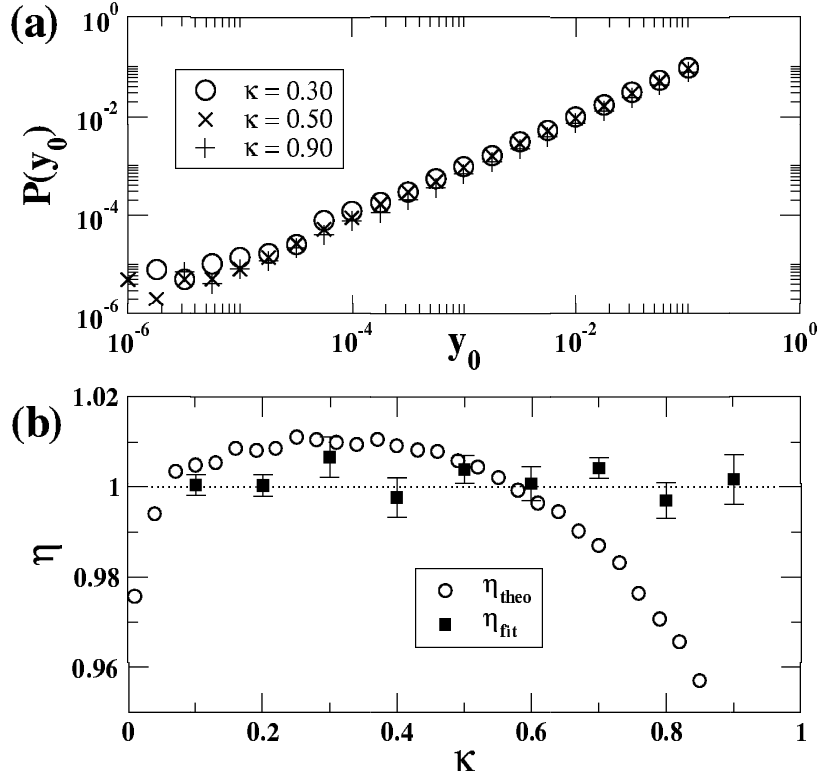

FIG. 5: (a) Fraction of points belonging to the basin of attractor B as a function of the distance to the attractor A, for different values of $\kappa$. (b) Variation of the scaling exponent with $\kappa$.

On the other hand, we argue that the scaling exponent $\eta$ is *exactly* equal to the unity, without invoking the stochastic model above. The argument lies on the symmetry of both attractors A and B with respect to their riddling properties, and holds only under these circumstances. Let $y_i$ be the transversal distance to the attractor at $I_i$, $i = 0, 1$, and $P_j(y_i)$ to be the probability of a trajectory, starting from a distance $y_i$ from the attractor at $I_i$, to belong to the basin of the attractor at $I_j$, $j \neq i$. Since the distance between the attractors is equal to the unity, $y_0 = 1 - y_1$.

Now we use the fact that the basins are intermingled, and that there are no other basins in the system, in such a way that

$$P_1(y_0) + P_0(y_1) = P_1(y_0) + P_0(1 - y_0) = 1.$$

If we set $P_j(y_i) \sim y_i^\eta$ there follows that

$$y_0^\eta + (1 - y_0)^\eta = 1$$

has the only solution $\eta = 1$ for arbitrary $y_0$.

## B. Uncertain fraction of the initial conditions

The scaling law (26) conveys information on the measure of the basins of both attractors, but not about the how riddling does occur in arbitrarily fine scales [10]. In fact, the riddled basin of the chaotic attractor A is a fat fractal, i.e. a fractal set with positive Lebesgue measure, whose fine scale structure can be characterized by the so-called uncertainty exponent [33]. Let us consider again the line $y = y_0$ and choose randomly an initial condition on that line. We now choose randomly another initial condition with uniform probability within an interval of length $2\delta$ and centered at the first initial condition. If both points belong to different basins, they can be referred to as $\delta$-uncertain.

The fraction of $\delta$-uncertain points, denoted by $<p>$, may be interpreted as the probability of making a mistake when attempting to predict which basin the initial condition is in, given a measurement uncertainty $\delta$, and is expected to scale with the latter as

$$<p> \sim \delta^\phi, \tag{28}$$

where $\phi \geq 0$ is the corresponding scaling exponent. We expect, from general grounds, that for riddled basins $\phi$ takes on typically small values. In the extreme case of $\phi = 0$ the uncertain fraction becomes constant, so that no decrease in the uncertain fraction could be achieved regardless of any improvement in the accuracy with which the initial condition is determined. More commonly for riddled basins even a dramatic increase in the accuracy (e.g., many orders of magnitude) has a slight effect in reducing the final-state uncertainty.

Figure 6(a) depicts our numerical result for making the above numerical experiment for $y_0 = 0.50$ and different values of $\kappa$, confirming the power-law nature of the scaling of $<p>$ *versus* the uncertainty radius $\delta$. The scaling exponent depends on the parameter $\kappa$, as shown in Fig. 6(b), and assumes values typically small, of the order of $10^{-2}$ or even less.

To show the consequences of having a so small value of $\phi$, let us suppose for example that $\phi = 10^{-2}$. If we managed to improve in such a large extent the accuracy with which the initial condition is determined that the uncertainty radius decrease by a factor of 100, the corresponding decrease in the uncertain fraction (or the probability of making a mistake) is only of

$$\frac{|<p'> - <p>|}{<p>} \times 100\% \approx 5\%,$$

in spite of the huge effort employed on improve the accuracy of measure. This is particularly disappointing when considering the experiments on the populations of *Tribolium* mentioned in the Introduction.

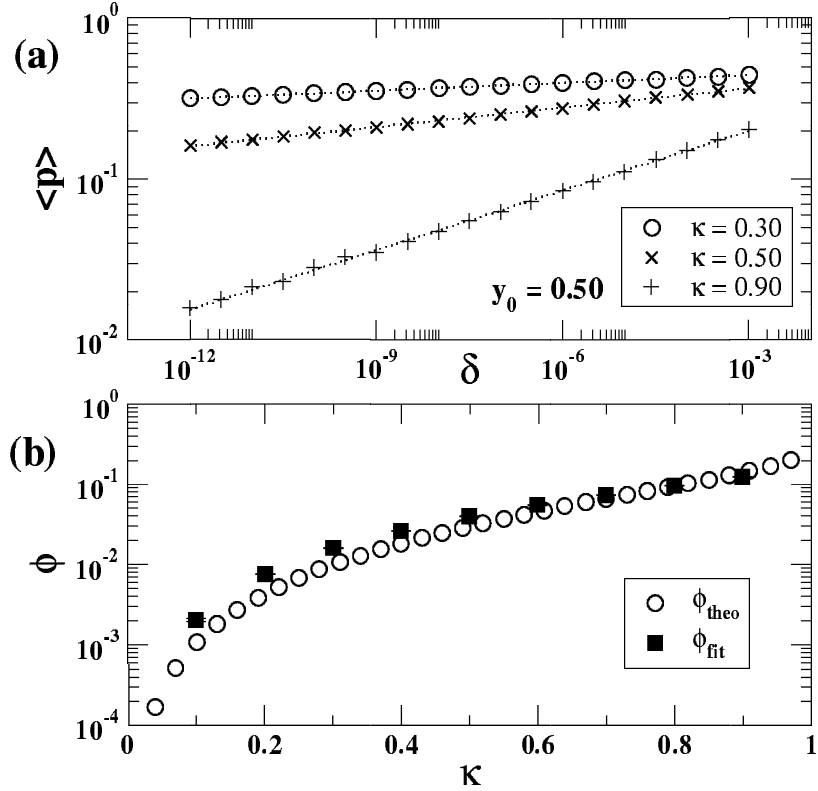

FIG. 6: (a) Fraction of $\delta$-uncertain initial condition *versus* radius of uncertainty $\delta$ for different values of $\kappa$. (b) Variation of the scaling exponent with $\kappa$

The stochastic model of Ott *et al.* gives for this exponent the following expression in terms of the infinite-time Lyapunov exponents [10, 17]

$$\phi = \frac{\lambda_\perp^2}{4D\lambda_\parallel}. \tag{29}$$

The results predicted by this expression are compared in Fig. 6(b) with the numerical results, and the agreement is quite good for all values of $\kappa$.

## VI. CONCLUSIONS

Intermingled basins, when occurring in a chaotic dynamical system, present many challenges for theoretical and experimental investigation. The extreme sensitivity to initial conditions make those systems highly susceptible to uncertainties of parameter and state determination. On the other hand, the rather stringent mathematical conditions necessary for the existence of intermingled basins (namely, the invariant manifold structure related to each coexisting attractor) limit the horizon of dynamical systems to be investigated. For example, coupled chaotic systems commonly present one such invariant manifold - the synchronization subspace. However, other invariant subspaces would require additional symmetry properties that not all coupled systems are able to exhibit.

One of the relatively few systems possessing these exceptional characteristics is the problem of competition of two species of *Tribolium sp.*, for which the two attractors represent extinction of either species. We have analyzed one model belonging to a more general class proposed by Hofbauer and collaborators [6], for which intermingled basins are known to exist for a wide interval of parameters. Our aim was to develop a series of topological and metric characterizations of intermingled basins that may be used in other dynamical systems for which the existence of those basins is not clearly demonstrated. This applies, for example, to coupled circuits and other high-dimensional chaotic systems where intermingled basins have been suggested to occur [18].

Two approaches were used to characterize intermingled basins in the two-species competition model. Firstly we investigated the mathematical conditions for intermingled basins through a topological and metric characterization of orbits in the coexisting attractors. We can use either typical chaotic orbits or atypical unstable periodic orbits embedded in the chaotic attractors. The methodology used in both cases is different: for typical orbits we rely on the properties of the transversal finite-time Lyapunov exponents, whereas atypical trajectories are treated using periodic-orbit analysis, which considers the contribution to the natural measure produced by unstable orbits of different periods. We expect that, if the period is large enough, the results of both procedures converge. We verified this concordance for the moments of the probability distribution function of the finite-time transversal Lyapunov exponent. Moreover, the contrast measure that compares the contribution of transversely stable and unstable periodic orbits reinforce the conclusions obtained with typical chaotic orbits.

A by-product of our analysis is that, besides riddling the *Tribolium* system also exhibits (for all values of the characteristic parameter $\kappa$) a strong form of non-hyperbolic behavior called unstable dimension variability, that means the coexistence (in the chaotic attractors) of periodic orbits with a different number of unstable directions. The presence of unstable dimension variability leads to unavoidable and severe problems of shadowability of chaotic trajectories. In particular, the shadowing time may be so small (when $\kappa$ goes to zero) that individual chaotic trajectories may not be useful for making predictions, even in the short run, since one-step errors are uncontrollably amplified by the non-hyperbolic dynamics.

The application of the methodology just described, specially periodic-orbit analysis, requires some detailed knowledge of the governing equations yielding the coexisting attractors. However, when only information on the basin structure is available, a better characterization of intermingled basins would be furnished by scaling laws. Two of them are considered in this paper: the fraction of the basin area of some attractor in the neighborhood of the other one, and the fraction of uncertain initial condition.

We used a general argument for proving that the first scaling exponent (relating the fraction of basin area) is equal to the unity, using only the symmetry properties of the system we deal with. This result has been confirmed by our numerical experiments and also compared with the theoretical prediction of a stochastic model proposed by Ott and collaborators [17]. The latter agrees best with the exact result when the nonlinearity parameter $\kappa$ approaches the value ($\kappa = 0$) for which the unstable dimension variability is the most intense. This can be explained by the nearly equal contribution of transversely stable and unstable orbits in this case. Since the transverse finite-time exponent quantify the average rate of shrinking (for stable) or expansion (for unstable) orbits, an approximately equal contribution of both kinds of orbits makes the statistical behavior of Lyapunov exponents more akin to a random walk, as assumed by the theoretical model of Ref. [17].

The second scaling law (uncertain fraction of initial condition) gives us the probability of making a wrong prediction about the future behavior of the system, or to what attractor will the system asymptote to. Our numerical results point to a value near zero (more precisely a value between $10^{-3}$ and $10^{-1}$), depending on the $\kappa$ parameter, in good accordance with the statistical model of Ref. [17], specially in the $\kappa \to 0$ limit where the exponent vanishes rapidly. The smallness of the scaling exponent, in this case, indicates that even a big improvement in the accuracy of the initial condition has little effect on our ability to predict the eventual outcome of the system (i.e., what species will become extinct). This explains, at least from the formal point of view, why in laboratory experiments with *Tribolium*, the species to become extinct was observed to change very sensitively with the specification of initial conditions and environmental conditions. Similar observations could be made for similar dynamical systems of physical and biological interest which are thought to present intermingled basins.

**Acknowledgments**

This work was partially supported by CNPq, CAPES, and Fundação Araucária (Brazilian government agencies). S. E. S. P. would like to acknowledge Dr. A. M. Saleh for presenting this problem to him.